# A Biologically Inspired Model of Distributed Online Communication Supporting Efficient Search and Diffusion of Innovation

**Soumya Banerjee**[1, 2, 3]

[1]Broad Institute of MIT and Harvard
 Cambridge, USA

[2]Ronin Institute
 Montclair, USA

[3]Complex Biological Systems Alliance
 North Andover, USA

## ABSTRACT

We inhabit a world that is not only "small" but supports efficient decentralized search – an individual using local information can establish a line of communication with another completely unknown individual. Here we augment a hierarchical social network model with communication between and within communities. We argue that organization into communities would decrease overall decentralized search times. We take inspiration from the biological immune system which organizes search for pathogens in a hybrid modular strategy.

Our strategy has relevance in search for rare amounts of information in online social networks and could have implications for massively distributed search challenges. Our work also has implications for design of efficient online networks that could have an impact on networks of human collaboration, scientific collaboration and networks used in targeted manhunts.

Real world systems, like online social networks, have high associated delays for long-distance links, since they are built on top of physical networks. Such systems have been shown to densify i.e. the average number of neighbours that an individual has increases with time. Hence such networks will have a communication cost due to space and the requirement of building and maintaining and increasing number of connections. We have incorporated such a non-spatial cost to communication in order to introduce the realism of individuals communicating within communities, which we call participation cost.

We introduce the notion of a community size that increases with the size of the system, which is shown to reduce the time to search for information in networks. Our final strategy balances search times and participation costs and is shown to decrease time to find information in decentralized search in online social networks. Our strategy also balances strong-ties (within communities) and weak-ties over long distances (between communities that bring in diverse ideas) and may ultimately lead to more productive and innovative networks of human communication and enterprise. We hope that this work will lay the foundation for strategies aimed at producing global scale human interaction networks that are sustainable and lead to a more networked, diverse and prosperous society.

*Corresponding author, : soumya.banerjee@roninstitute.org;

## KEY WORDS

Social computing, complex systems, social dynamics, innovation diffusion, artificial immune system

## CLASSIFICATION

ACM

## INTRODUCTION

We inhabit a world that is not only "small" but supports efficient decentralized search – an individual using local information can establish a line of communication with another completely unknown individual. This property of "six degrees of separation" had been uncovered by the celebrated work of Stanley Milgram [1]. In his social experiment, individuals were instructed to forward letters to completely unknown persons using only information about their immediate acquaintances. Milgram found that contrary to intuition despite the immense geographical distances separating the source and target individuals, a line of communication could nevertheless be established between individuals separated by great social and geographic distances. Mathematical models have been recently proposed to explain this phenomenon [2-5]. Drawing inspiration from biological systems, we extend the rich body of understanding that has grown around this puzzling social phenomenon by adding several levels of realism into these models.

Kleinberg [2-4] proposed that the presence of long-distance links obeying a precise probability distribution facilitated decentralized search by laying latent structural "cues" in the underlying network to guide this local search process towards a global target. We extend his model by making the observation that social networks are divided into communities with communication between and within communities. Previous models have also ignored the effect of physical space on decentralized search times. Individuals also incur costs upon establishing and maintaining social contacts [6]. Current work has ignored such non-spatial costs and we remedy that by introducing a "social participation cost". Our final contribution is to examine the role of physical space and innovations in communication technologies upon search times in social networks.

In our work, we draw inspiration from a biological system which is optimized for decentralized search – the immune system (IS). The IS is tasked with finding small quantities of pathogens while searching through the physical space of the whole body. Additionally the site of infection has to recruit IS cells over large physical distances. Hence search in the IS is constrained by physical space of the body. Current work in social network models implicitly ignore considerations of space [2-4], since long-distance links are assumed to have the same cost as short-distance links. Real world systems, like online social networks, however lie somewhere along this continuum of physical space, since they are built on top of physical networks like the World Wide Web and infrastructure networks like power grids.

To efficiently solve the difficult search process described above, the IS is organized into modular structures called lymph nodes (LN). The search for small amounts of pathogen is now modular – a small search through the physical space of the LN and parallel – many such small searches in parallel. In the context of social networks, individuals have been observed to mix assortatively i.e. they make more connections with individuals with whom they share characteristics. Our key inspiration from the immune system is to recognize that search can be efficient by organizing the search space into "communities", where search within a community could be faster than search between communities.

The remainder of the paper is organized as follows – we present a brief overview of the immune system and how it has evolved to conduct efficient search (Section Immunological Preliminaries); we review current work on efficient decentralized search in social networks (Section Hierarchical Social Network Model and Section Decentralized Search) and then we extend the models (Section A Mathematical Model of Social Networks Incorporating Community Communication) by

    a) incorporating communication between and within communities,
    b) adding a "social participation cost", and

    c) introducing the effect of physical space and innovations in communication technologies as a variable model parameter (Section Role of Technology in Inter-Community Communication)

We analyze the model and show that it supports even faster decentralized search than exhibited by current models of social networks. Finally we make concluding remarks and outline future work.

## Immunological Preliminaries

The area of tissue that drains into a lymph node (LN) is called its draining region. Immune system cells called dendritic cells sample the tissue in the draining region (DR) for pathogens, and upon encountering them, migrate to the nearest LN and signal to other immune system cells called B-cells to secrete chemicals called antibodies.

A horse 25,000 times larger than a mouse must generate 25,000 times more absolute quantities of antibody in order to achieve the same concentration of antibody in the blood (where blood volume is $\propto$ M (host body mass) [27]). A fixed antibody concentration is required to fight infections.

If organisms of all body sizes activated the same number of B cells, the time for a fixed number of B cells to produce antibody Ab is $\propto \log_2 M$ (since immune system cells reproduce exponentially through clonal amplification). For example, since it takes 4 days of exponential growth of activated B cells to produce sufficient anti-WNV neutralizing antibody in mice [28], then the corresponding time for a horse would be more than 2 months. This conflicts with empirical data on horses [7]. We assume that the immune system of larger organisms has to activate a number of antigen-specific B cells $\propto$ M, in order to build up the critical density of antibodies in a fixed period of time.

Since only a very small number of pathogen-specific B cells reside inside the LN (1 in $10^6$ IS cells) [7], this implies that as the organism size increases, the infected site LN has to recruit increasing numbers of B cells from other LNs. However having a larger LN means that the volume of the DR it services will be very large (since the total amount of lymphoid tissue is proportional to body mass), which will increase the average time taken by immune system cells to reach the LN. Hence we see that there is a tradeoff between having a larger LN (reduces global communication time involved in recruiting additional immune system cells to ensure a global antibody response) and a smaller LN (reduces local communication time involved in trafficking antigen-loaded dendritic cells to the draining LN).

In previous work [7] we showed that there is a tradeoff between local intra-node communication (detecting antigen within a node) and global inter-node communication (efficiently distributing antibody throughout the system). Having a huge node with all immune system cells within it would naturally minimize the global communication time but increase the local antigen detection time. On the contrary, having numerous small nodes would minimize the local communication time associated with detecting antigen but only at the expense of increased global communication time between the nodes. The optimal architecture that balances these two opposing goals is one where nodes become "larger" as the system size increases (to minimize global communication costs) as well as more numerous (so as to minimize the search volume within a node and hence the associated local communication cost)[7-14].

Similar tradeoffs between local and global communication times exist for other system. We have previously applied such immune system inspired strategies to human-engineered systems like mobile robots, mobile networks, peer-to-peer systems [7-14], distributed systems of computer networks and process scheduling [15], automated program verification [18] and also dynamics of crime in human societies [16]. Here we extend these models to social networks.

## A Hierarchical Social Network Model

Social structures exhibit self-similarity, with individuals organizing their social contacts hierarchically. Pairs of individuals belonging to the same small community form stronger social ties than pairs of individuals who are related by membership in a larger community. Follow-up work on the Milgram experiment [2,3] found that participant decisions were facilitated by geographical and occupational clues. The hierarchical model incorporates a community structure based on a social attribute like occupation and makes concrete the concept of a social distance. Such recursive community structures are also present in:

a) the World Wide Web, where a music file can be classified as an operatic composition, or more generally about music and still more generally about arts,

b) Computer networks like Autonomous Systems form tight groups based on geography, which are in turn composed of even tighter communities.

c) Citation networks are constituted of compact groups based on subject (like Physics and Chemistry) and even smaller groups based on further subject classifications (like Astronomy and Inorganic Chemistry).

Unifying these concepts and making these notions concrete, Kleinberg et al. [2] represented the structure of communities within communities as a perfectly balanced tree $T$ with constant fanout $b$ and height $h$. The number of leaf nodes is $n = b^h$. We note that in this model the fanout $b$ is fixed and does not change as the size of the system or total population increases. The social distance between two nodes is defined as the height of their lowest common ancestor. A random graph $G$ was then constructed using the leaf nodes as vertices. Assuming that the density of link pairs between nodes is lower for nodes that have a higher social distance, the authors settled on the following form of the probability of a link between two nodes

$$\Pr[e(u,v)] \propto b^{-\beta h(u,v)}$$

where $e(u,v)$ denotes an edge between leaf nodes $u$ and $v$, $\beta$ is a model parameter and $h(u,v)$ is the social distance between the nodes (height of the lowest common ancestor of the nodes $u$ and $v$).

## Decentralized Search

Kleinberg et al. investigated the behaviour of this model with respect to how it facilitated decentralized search i.e. search without global knowledge where each node only has information about its neighbours (a formalization of the Milgram experiment [1] where each participant was devoid of global knowledge of the system and was endowed with only local knowledge about their acquaintances). They found that a decentralized algorithm can achieve the best performance when the model parameter $\beta = 1$. Any higher value of $\beta$ (which would decay the probability of long-distance links faster) or a lower value (which would lead to an almost uniform distribution of long and short distance links) would degrade performance. The intuition behind this is that with an uniform distribution of long and short distance links, any node is faced with a disorienting array of choices form which to choose the next node to forward the message to. With the precise distribution of links, the network structure itself provides latent structural cues, such that each node with only local information can guide a message to a distant target.

Further, with a node out-degree $k = c(\log n)^2$, the expected delivery time was proven to be $O(\log n)$. These results are in agreement with Milgram's original experiments [1] and lend insight into the puzzling riddle of "six degree's of separation" in social networks. Analysis of empirical data on social networks by Adamic et al. on email networks and Liben-Nowell et al. [19] on a social networking website reveal that such networks closely approximate the property of efficient search predicted by this simple model.

## A Mathematical Model of Social Networks Incorporating Community Communication

Thinking in terms of social networks, lymph nodes are constituted of increasingly larger aggregations of IS cells. Such "communities" of IS cells grow bigger in order to compensate for increasing costs to communicating with other communities [7]. However there are increasing local communication costs incurred by a larger community and an efficient architecture will balance the needs of both local and global communication.

Previous work in social networks has not considered clustering of individuals into communities, leading to faster communication within communities as opposed to communication between communities. Here we augment the hierarchical social network model with communication between and within communities. We argue that organization into communities would decrease overall search times.

Social networks exhibit strong community structure. Such communities typically confer upon members advantages of efficient communication and collaboration. Presence of larger communities would mean a reduced number of total communities. Since global communication is between communities and each community has to interact with fewer communities, this would also reduce global search time. Communities however could have significant communication costs within themselves and we model this cost generally allowing it to take any value depending on a model parameter. Communities also introduce a "participation cost" – an individual link to a neighbour introduces costs of building and maintaining communication [6].

In other words, having large community sizes decreases global communication time. However this comes at the cost of increased local communication time and a social participation cost. We now make this notion explicit by a mathematical model and determine the precise size of a community, as a function of the number of individuals, that would balance the opposing tradeoffs of global communication on one hand, and local communication and participation cost on the other.

We extend the previous hierarchical social network model (summarized in Section A Hierarchical Social Network Model) by incorporating communication between and within communities. We have a perfectly balanced tree $T$ with constant fanout $b$ and height $h$. The number of individuals and leaf nodes is $n = b^h$. A community is defined as a group of leaf nodes which have the height of their lowest common ancestor as $h-1$ and hence the size of a community is $b$. The social distance between two communities is defined as the height of their lowest common ancestor. A random graph $G$ is then constructed by treating communities as vertices. The number of such vertices and communities is $n/b$. Analogously to Kleinberg et al [2] we assume that the density of link pairs between communities is lower for communities that have a higher social distance, and settled on the following form of the probability of a link between two communities

$$\Pr[e(c_1,c_2)] \propto b^{-h(c_1,c_2)}$$

where $e(c_1,c_2)$ denotes an edge between communities $c_1$ and $c_2$ and $h(c_1,c_2)$ is the social distance between the communities (height of the lowest common ancestor of the communities $c_1$ and $c_2$).

We model the connections within a community as forming a clique with $b$ individuals. Later we relax this assumption. The local participation cost of maintaining these connections is given

$$c_{local} = \kappa_1 n(b-1)$$

since each individual has $b-1$ local connections within the community. The number of connections that a node must maintain in order to ensure a polylogarithmic gossip time is given by $O((\log x)^2)$ where $x$ is the number of nodes [2,3]. Since the number of communities is given by $n/b$, the global participation cost for all communities is given by

$$c_{global} = \kappa_2 \frac{n}{b}(\log(n/b)^2)$$

The time required for local communication between individuals in a community is now a constant since we model the community as a completely connected graph

$$t_{local} = \kappa_3$$

Lastly, the time required for global communication between $x$ individuals is given by $O(\log_b x)$ [2]. The corresponding time for $n/b$ individuals is given by

$$t_{global} = \kappa_4 \log_b(n/b)$$

We intend to minimize both the social participation cost and the communication time. We emphasize both the quantities equally by minimizing the logarithm of the costs and the time, thus ensuring that the number of individuals enters into both terms logarithmically.

$$X = \log c_{local} + \log c_{global} + t_{local} + t_{global}$$

Simplifying we have

$$X = \log \kappa_1 + \log \kappa_2 + 2\log n + 2\log\log(n/b) + \kappa_3 + \kappa_4 \frac{\log n}{\log b} - \kappa_4$$

Differentiating with respect to $b$ and simplifying further we have

$$(\log b)^2 b = \kappa_4 \log n$$

$$b = O((\log n)^\varepsilon), \text{for some } \varepsilon < 1$$

We note that in the original model by Kleinberg [2] the fanout $b$ is fixed and does not change with the size of the system. In contrast in our model $b$ will change as the size of the system or total population changes. Substituting the scaling relation for $b$ in the equation for communication time we have (the derivations are outlined in greater detail in the Appendix section)

$$t_{local} + t_{global} = \kappa_3 + \kappa_4 \log_b(n / b)$$

$$= \kappa_3 + \kappa_4 \frac{\log n}{\log b} - \kappa_4$$

$$= O\left(\frac{\log n}{\log \log n}\right)$$

This is a clear improvement on the current best time for decentralized gossip of $O(\log_b n)$ [2,3]. The optimal strategy is a hybrid modular strategy where the size of communities and the number of communities both increase with the size of the system (shown in Fig. 1). Such a strategy would decrease the time required to search for rare information in online social networks, co-ordinated manhunts, or facilitate information exchange and discovery in scientific collaboration networks [17]. We also note that in this strategy the number of communities (given by *n/b*) would also increase as

$$O\left(\frac{n}{(\log n)^\varepsilon}\right), \text{for some } \varepsilon < 1$$

The preceding analysis clearly demonstrates that increasing community size as a function of system size can decrease time for decentralized search and gossip, whilst keeping social participation costs within manageable levels. In the next section, we show that our model predictions remain invariant even if we relax one of our assumptions.

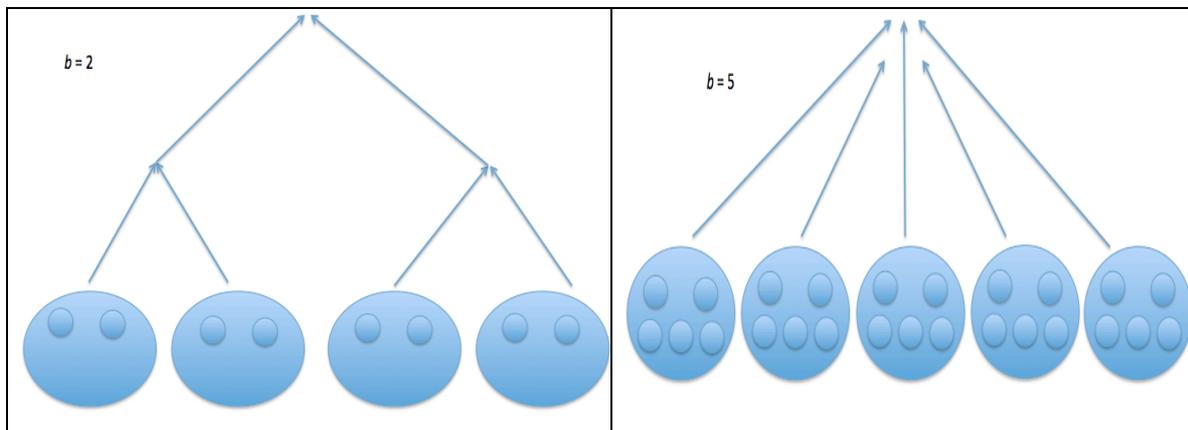

**Figure 1.** A simplified depiction of our hybrid modular architecture. Left Panel: interaction network with 2 persons per community and a branching factor of 2. Right Panel: A scaled up version of the above network with 5 persons per community and a branching factor of 5 (both the number of communities and the number of people in a community increases).

# Role of Technology in Inter-Community Communication

Technological advances have enabled faster modes of communication. Newer advances in communication (like fibre-optic technologies) have an impact on communication times between communities. Future communication technologies have the potential to radically alter the cost of communication. Here we relax the assumption of a completely connected topology at the community level and instead model the local communication time as a general variable which is allowed to take on values depending on the level of sophistication of communication technologies. We still keep the same identity for local participation cost since it is a conservative estimate but change the local communication time as follows

$$t_{local} = \kappa_3 b^{\omega}$$

where $\omega = 0$ represents an advanced technology that imposes constant communication overhead and $\omega = 1$ corresponds to a technology that constrains communication within communities to proceed in a serial manner.

Now we have the following equation to be minimized

$$X = \log c_{local} + \log c_{global} + t_{local} + t_{global}$$

$$= \log \kappa_1 + \log \kappa_2 + 2\log n + 2\log \log(n/b) + \kappa_3 b^{\omega} + \kappa_4 \frac{\log n}{\log b} - \kappa_4$$

Differentiating with respect to $b$ and upon simplification we have

$$(\log b)^2 \left[ \omega b^{\omega} - \frac{2}{\log(n/b)} \right] = \kappa_4 \log n$$

Noting that in the limit, the expression within parentheses tends to $\omega b^{\omega}$, we have

$$\omega b^{\omega} (\log b)^2 = \kappa_4 \log n$$

$$b = O((\log n)^{\varepsilon}), \text{for some } \varepsilon < 1/\omega$$

Substituting this identity for $b$ in the relation for communication time we have (the derivations are outlined in greater detail in the Appendix section)

$$t_{local} + t_{global} = \kappa_3 + \kappa_4 \log_b(n/b)$$

$$= O\left( \frac{\log n}{\log \log n} \right)$$

This is just a reiteration of our previous result improving on the previous *O(logn)* bound on gossip time. Thus relaxing our assumptions about the communication topology or technology within communities, does not change the nature of our model behaviour.

In summary our strategy balances participation cost and search time and leads to optimal search time that is faster than the current best time for decentralized gossip of $O(\log_b n)$ [2]. The optimal strategy is a hybrid modular strategy where the size of communities and the number of communities both increase with the size of the system (shown in Fig. 1).

## Discussion

The results discussed by Kleinberg [2-4] implicitly ignore considerations of space, since long-distance links are assumed to have the same cost as short-distance links. Work in the immune system, especially investigations into how LN size and numbers scale with body size, is motivated primarily by constraints of space. An infected site LN has to incur increasing communication costs in larger organisms, since it has to recruit IS cells over larger physical distances. This is clearly a case where long-distance links have higher associated costs. Real world systems, like online social networks, however lie somewhere along this space continuum, since they are built on top of physical networks like the World Wide Web and infrastructure networks like power grids. Such systems have been shown to densify i.e. the average number of neighbours that an individual has increases with time [20]. Hence such networks will have a communication cost not wholly attributable to space but due to the requirement of building and maintaining and increasing number of connections. We have incorporated such a non-spatial cost to communication in order to introduce the realism of individuals communicating within communities.

This work is inspired by conceptual similarities between a lymph node and a social community. However we would like to point out that there are some technical differences that exist between the two, which nevertheless do not change the outcome of our model predictions. A lymph node is a very diverse community, being composed of populations of immune system cells with very different specificities to pathogens. In contrast, social networks exhibit assortative mixing where individuals aggregate with others based on attribute similarity.

Milgram's social connectedness experiment continues to have relevance in the context of online social networks. Many explanations have been proposed to explain how completely unrelated individuals could communicate with each other based on just information about immediate acquaintances. Kleinberg [2-4] proposed that the presence of long-distance links obeying a precise probability distribution facilitated decentralized search by laying latent structural "cues" in the underlying network to guide this local search process towards a global target. Watts et al. [5] proposed that decentralized search is facilitated by individuals searching for multiple attributes e.g. individuals have associated geographical attribute as well as an occupation.

Our work extends previous social network models by incorporating the additional realisms of aggregation of individuals into communities, communication between and within communities, and a social participation cost. Communities confer benefits of efficient communication and collaboration upon members. Presence of larger communities would mean a reduced number of total communities. Since global communication is between communities and each community has to interact with fewer communities, this would reduce global search time. Communication cost within communities is modeled generally and allowed to take on values depending on network topology and sophistication of communication technology. All the advantages of communities however come at the expense of a "participation cost" – an individual link to a neighbour introduces costs of building and maintaining communication [6]. We note the caveat that the size of a community may not increase indefinitely and could be limited by the so-called Dunbar limit on the number of connections that can be sustained by the human brain.

Our model demonstrates that increasing community sizes would lead to efficient decentralized search and gossip in social networks while minimizing social participation costs. Here we augment the hierarchical social network model with communication between and within communities. We argue that organization into communities would decrease overall search times.

Our final strategy is inspired by similar strategies employed by the immune system [7]. Our results are relevant for search for rare amounts of information in online social networks as exemplified by the DARPA Balloon Challenge where participants had to find 3 balloons spread throughout the US with only local information (using only their friends and immediate acquaintances).

Our work also has potential implications for design of efficient online networks that could have an impact on networks of human collaboration, scientific collaboration and networks used in targeted manhunts. It may be possible to design virtual "lymph nodes" or communities whose size would need to scale with the system size. Within each community search would be very fast and whenever they would need to find information from other communities they would reach out to them (the system is diagrammed in Fig. 1).

We note that our proposed strategy shows characteristics of densification which has the additional benefit of enabling faster and more efficient diffusion of innovation and may lead to more efficient economic generation activities as proposed in earlier models in cities [21-24].

The results discussed by Kleinberg [2-4] implicitly ignore considerations of space, since long-distance links are assumed to have the same cost as short-distance links. Work in the IS, especially investigations into how LN size and numbers scale with body size, are motivated primarily by constraints of space. An infected site LN has to incur increasing communication costs in larger organisms, since it has to recruit immune system cells over larger physical distances. This is clearly a case where long-distance links have higher associated costs. Real world systems, like online social networks, however lie somewhere along this space continuum, since they are built on top of physical networks like the World Wide Web and infrastructure networks like power grids. Such systems have been shown to densify i.e. the average number of neighbours that an individual has increases with time [20]. Hence such networks will have a communication cost not wholly attributable to space but due to the requirement of building and maintaining and increasing number of connections. We have incorporated such a non-spatial cost to communication in order to introduce the realism of individuals communicating within communities.

Large human agglomerations have disproportionately more wealth [24] and this is attributed to the increasing density of productive human networks that generate wealth. In larger human agglomerations and cities there are more people and this means there are more choices for establishing relationships [21-23]. This also decreases the time for ideas and innovations to diffuse.

However number of connections cannot increase without imposing a cost. We optimize both the number of connections (and hence productivity) and also search times. Hence our final strategy optimizes both innovation by densification of the social interaction network [23] and the time to find rare information. It achieves this by fundamentally maximizing the rate of information diffusion. Granovetter has argued about the strength of weak ties [25]. Our approach balances strong-ties (within communities) and weak-ties over long distances (between communities that bring in diverse ideas) and may ultimately lead to more productive and innovative networks of human communication and enterprise [17].

We hope that this work will lay the foundation for strategies aimed at producing global scale human interaction networks that are sustainable and lead to a more networked, diverse and prosperous society.

**Acknowledgements** The author wishes to acknowledge fruitful discussions with Dr. Melanie Moses.

## Appendix

Here we outline some of the derivations in greater detail.

### A Mathematical Model of Social Networks Incorporating Community Communication

We intend to minimize both the social participation cost and the communication time. We emphasize both the quantities equally by minimizing the logarithm of the costs and the time, thus ensuring that the number of individuals enters into both terms logarithmically.

$$X = \log c_{local} + \log c_{global} + t_{local} + t_{global}$$

Simplifying we have

$$X = \log(\kappa_1 n(b-1)) + \log(\kappa_2 \frac{n}{b}[\log(n/b)]^2) + \kappa_3 + \kappa_4 \frac{\log n}{\log b} - \kappa_4$$

$$= \log \kappa_1 + \log n + \log(b-1) + \log \kappa_2 + \log(n/b) + \log[\log(n/b)]^2 + \kappa_3 + \kappa_4 \frac{\log n}{\log b} - \kappa_4$$

$$= \log \kappa_1 + \log \kappa_2 + 2\log n + 2\log\log(n/b) + \kappa_3 + \kappa_4 \frac{\log n}{\log b} - \kappa_4$$

Differentiating with respect to $b$ we have

$$\frac{dX}{db} = 0 + \frac{2}{\log(n/b)} \frac{1}{n/b} \frac{-n}{b^2} + 0 - \kappa_4 \frac{\log n}{(\log b)^2 b} - 0 = 0$$

Simplifying further we have

$$1 = \frac{2}{b\log(n/b)} + \frac{\kappa_4 \log n}{b(\log b)^2}$$

$$b(\log b)^2 = \frac{2(\log b)^2}{\log(n/b)} + \kappa_4 \log n$$

$$(\log b)^2 \left[ b - \frac{2}{\log(n/b)} \right] = \kappa_4 \log n$$

Noting that the expression within parentheses is going to tend towards b, we have

$$(\log b)^2 b = \kappa_4 \log n$$

$$b = O((\log n)^\epsilon), \text{for some } \epsilon < 1$$

Substituting the scaling relation for *b* in the equation for communication time we have

$$t_{local} + t_{global} = \kappa_3 + \kappa_4 \log_b(n/b)$$

$$= \kappa_3 + \kappa_4 \frac{\log n}{\log b} - \kappa_4$$

$$= \kappa_3 + \kappa_4 \frac{\log n}{\log(d(\log n)^\varepsilon)} - \kappa_4, \text{where } d \text{ is a constant}$$

$$= O\left(\frac{\log n}{\log d + \log \log n}\right)$$

$$= O\left(\frac{\log n}{\log \log n}\right)$$

## Role of Technology in Inter-Community Communication

Here we derive an expression for decentralized search time and the effect of technology. We keep the same identity for local participation cost since it is a conservative estimate but change the local communication time as follows

$$t_{local} = \kappa_3 b^\omega$$

where $\omega = 0$ represents an advanced technology that imposes constant communication overhead and $\omega = 1$ corresponds to a technology that constrains communication within communities to proceed in a serial manner.

Now we have the following equation to be minimized

$$X = \log c_{local} + \log c_{global} + t_{local} + t_{global}$$

$$= \log(\kappa_1 n(b-1)) + \log(\kappa_2 \frac{n}{b}[\log(n/b)]^2) + \kappa_3 b^\omega + \kappa_4 \frac{\log n}{\log b} - \kappa_4$$

$$= \log \kappa_1 + \log n + \log(b-1) + \log \kappa_2 + \log(n/b) + \log[\log(n/b)]^2 + \kappa_3 b^\omega + \kappa_4 \frac{\log n}{\log b} - \kappa_4$$

$$= \log \kappa_1 + \log \kappa_2 + 2\log n + 2\log \log(n/b) + \kappa_3 b^\omega + \kappa_4 \frac{\log n}{\log b} - \kappa_4$$

Differentiating with respect to $b$ we have

$$\frac{dX}{db} = 0 + \frac{2}{\log(n/b)} \frac{1}{n/b} \frac{-n}{b^2} + \kappa_3 \omega b^{\omega-1} - \kappa_4 \frac{\log n}{(\log b)^2 b} - 0 = 0$$

Upon simplification we have

$$\omega b^{\omega-1} = \frac{2}{b \log(n/b)} + \frac{\kappa_4 \log n}{b (\log b)^2}$$

$$\omega b^{\omega} (\log b)^2 = \frac{2(\log b)^2}{\log(n/b)} + \kappa_4 \log n$$

$$(\log b)^2 \left[ \omega b^{\omega} - \frac{2}{\log(n/b)} \right] = \kappa_4 \log n$$

Noting that in the limit, the expression within parentheses tends to $\omega b^{\omega}$, we have

$$\omega b^{\omega} (\log b)^2 = \kappa_4 \log n$$

$$b = O((\log n)^{\varepsilon}), \text{for some } \varepsilon < 1/\omega$$

Substituting this identity for b in the relation for communication time we have

$$t_{local} + t_{global} = \kappa_3 + \kappa_4 \log_b (n/b)$$

$$= \kappa_3 + \kappa_4 \frac{\log n}{\log b} - \kappa_4$$

$$= \kappa_3 + \kappa_4 \frac{\log n}{\log(d(\log n)^{\varepsilon})} - \kappa_4, \text{where } d \text{ is a constant}$$

$$= O\left( \frac{\log n}{\log d + \log \log n} \right)$$

$$= O\left( \frac{\log n}{\log \log n} \right)$$